\documentclass[twocolumn,showpacs,preprintnumbers,amsmath,amssymb]{revtex4}
\usepackage{graphicx}% Include figure files
\usepackage{dcolumn}% Align table columns on decimal point
\usepackage{bm}% bold math
\usepackage{epsfig}

\begin{document}

\preprint{APS/123-QED}

\title{Self-overlap as a method of analysis in Ising models}% Force line breaks with \\

\author{A Ferrera, B Luque, L Lacasa$^*$ and E Valero}
\email{lucas@dmae.upm.es}

\affiliation{Dpto. de Matem\'atica Aplicada y Estad\'istica\\
ETSI Aeron\'auticos\\ Universidad Polit\'ecnica de Madrid.
}%

\date{\today}% It is always \today, today,

\begin{abstract}
The damage spreading method (DS) provided a useful tool to obtain
analytical results of the thermodynamics and stability of the $2D$
Ising model --amongst many others--, but it suffered both from
ambiguities in its results and from large computational costs. In
this paper we propose an alternative method, the so called
self-overlap method, based on the study of correlation functions
measured at subsequent time steps as the system evolves towards its
equilibrium. Applying markovian and mean field approximations to a
$2D$ Ising system we obtain both analytical and numerical results on
the thermodynamics that agree with the expected behavior. We also
provide some analytical results on the stability of the system.
Since only a single replica of the system needs to be studied, this
method would seem to be free from the ambiguities that afflicted DS.
It also seems to be numerically more efficient and analytically
simpler.
\end{abstract}

%\keywords{self-organization, hierarchy}
\maketitle

\section{Introduction}
The damage spreading (DS) method \cite{DS0} is a remarkable tool
amongst the many ones developed in recent years in the effort to
understand the dynamics of cooperative systems. Very roughly
speaking the goal of the method is to study the stability of a
cooperative system under a small perturbation: if perturbations die
off after some time then the system must be in a stable, ordered
state; if small perturbations always get amplified however then the
system must be in a disordered, chaotic state. By studying how far
away the final states are from the initial ones given the initial
perturbation one can get information about the system such as, for
example, the Lyapunov exponents.

Of course there is a key aspect that differentiates cooperative
systems from classical dynamical systems. Namely, that in the former
case given the complexity of the systems under study we almost never
have at our disposal a detailed analytical solution to the equations
of motion in order to study the system's stability under
perturbations. Here is where the DS method comes in: its operational
side amounts to an algorithm designed to study how small
perturbations spread within the system by working in detail how each
of the system's components react to the changes. The method has been
applied to many different dynamical systems such as Ising systems
\cite{Vojta1, Vojta2, Vojta3, Vojta4, Vojta5, DS1, DS2, DS3},
Kauffman networks \cite{DS4, DS4bis, DS5}, spin glasses \cite{DS6,
DS62}, cellular automata \cite{DS8} amongst others, yielding in many
cases useful information about their evolution and stability.
Succinctly speaking, the algorithm analyzes the evolution of two
almost identical states of the system.  The damage (difference
between the two initial states) is specified as part of the initial
conditions. That is, on one side we have a specified state of the
system, and on the other we have a replica that only differs in a
small perturbation (the damage) from this original state. One then
fixes the stochastic evolution to be the same for each replica (in a
Monte Carlo simulation the method imposes the same random numbers at
each step of time on both copies for instance). As we let the two
copies evolve, the method analyzes their distance (Hamming distance)
as a function of time. Useful information about the system can then
be extracted from this,
not only numerically but in some cases also analytically. \\

However, as was shown in \cite{mhas}, \cite{ja} and \cite{DS1} (and
references therein) DS has been shown to be ill-defined in the sense
that different --and equally legitimate-- algorithmic
implementations of the same physical system's dynamics can yield
different DS properties. This ambiguity stems from the fact that
while the transfer matrix for the evolution of a single system is
completely determined by the one-point correlation functions
\cite{DS1}, the simultaneous evolution of two replicas however is
governed by a joint transfer matrix determined by two-point
correlation functions. For example, Glauber and both standard and
uncorrelated heat bath (HB) algorithms satisfy detailed balance with
respect to the same Hamiltonian. It follows that these three
different update rules generate the same equilibrium ensemble and
are therefore equally legitimate to mimic the evolution in time of
an Ising system coupled to a thermal reservoir. Accordingly, the
one-point correlation functions for the three cases coincide and the
corresponding transfer matrices for single systems are identical. On
the other hand the two-point functions for HB and Glauber dynamics
are different; hence damage evolves differently in either case (see
\cite{DS1} and references for a extended quantitative version of
this argument). As long as the results depend on the algorithm being
implemented, one can not assert that the results obtained from a
given DS analysis are conclusive and unambiguous.

This handicap is a major motivation in order to search for an
alternative method of stability analysis. Our goal in this paper
will be to propose a different approach to study the stability of
cooperative systems. By relying heavily on the above mentioned fact
that the evolution of a single system is determined only by the
one-point correlation functions we will try to eliminate some of
ambiguities found in the DS method.\\

In order to be specific, as a test case we will focus on the study
of a well-known type of system: Ising models. In \cite{Vojta1} Votja
tackled the $2D$ Glauber-Ising model via DS. He obtained results on
the thermodynamics (magnetization, ferromagnetic transition) and
stability (regular vs. chaotic behavior) of the model both
analytically and numerically. Due to the nature of the method
however (at every step of time we must keep account of the two
replicas), there is obvious room for improving the computational
efficiency. This is also the case in the analytical realm, where
accounting for the way in which at each step of time the differences
between the two copies may increase inevitably leads to lengthy
computations (as was shown in \cite{Vojta1, Vojta2, Vojta3, Vojta4,
Vojta5}). This on itself constitutes
a second motivation in order to search for alternative methods.\\

The method of analysis that we will propose here, so-called the
self-overlap method (SO) \cite{SO1, SO2, SO3}, has already been
successfully used in the study of the stability and critical points
of random Boolean networks, a system that is multi-component albeit
deterministic. In this work we will show that the method can also be
successfully applied to a stochastic system such as a spin network.
We will obtain analytical and numerical results on the $2D$
Glauber-Ising model that exactly match those yielded by DS. However,
contrary to DS, SO proceeds by handling only one replica and
analyzing its own evolution in time --using basically one-point
correlation functions at subsequent time steps for that task. The
computational costs are thus lower in SO than in DS. As we will see
the analytical calculations also become much simpler while yielding
the same results. Furthermore, and what is more important, the SO
method is free from the ambiguities that afflicted DS due to its use
of two replicas. This comes as a direct consequence of the already
mentioned fact that on a single replica it does not matter whether
one uses Glauber or HB dynamics since they posses the same one-point
correlation functions.
\\

We will follow the development applied by Vojta in \cite{Vojta1},
comparing in each case the results obtained using DS and our results
(using SO). The paper is organized as it follows: in section II  we
quickly introduce both the $2D$ Glauber-Ising model and SO. We then
apply the method to the $2D$ Glauber-Ising model in section III,
obtaining a system of equations (master equation) that describe the
dynamical evolution of the system. We discuss then how to apply a
mean-field approximation to the system, and compare it with the
methodology used by Vojta \cite{Vojta1}. In section IV we obtain an
analytical expression for the magnetization of the system in both
ferromagnetic/paramagnetic phases similar to that obtained by Vojta
\cite{Vojta1}. Numerical results are provided at this point in order
to validate the mean field approximation assumed in the analytical
development. Finally, in section V we provide some analytical and
numerical results on the stability of the model, showing that the
system is chaotic (disordered) in the paramagnetic phase.
Conclusions are presented in section VI.

%%%%%%%%%%%%%%%%%%%%%%%%%%%%%%%%%%%%%%%%%%%%%%%%%%%%%%%%%%%%%%%%
\section{Ising model, the damage spreading vs. self-overlap method}
\subsection{Glauber Ising model}
We will work with a kinetic Ising model, a lattice of $N$ spins,
$s_i \in \{+1,-1\}$, that follows Glauber dynamics. That is, at
every time step a lattice site $i$ is chosen at random. If the spin
value of site $i$ at time $t$ is given by $s_i(t)$, at time $t+1$ it
will be given by:
\begin{equation}
s_i(t+1)={\rm sgn}\Biggl [\phi(h_i(t))-{\frac 12}+s_i(t)\Biggl( \xi
_i(t)-{\frac 12}\Biggr) \Biggr],\;\;\;\;\;\;\;\; \label{s(t+1)}
\end{equation}
where $\xi _i(t)$ is a random number such that $\xi _i(t)\in [0,1)$.
The transition probability $\phi (h_i)$ is given by the usual Glauber expression:
\begin{equation}
\phi (h_i(t))={\frac{e^{h_i(t)/T}}{e^{h_i(t)/T}+e^{-h_i(t)/T}}}%
\;\;\;\;\;\;\;\;  \label{fi}
\end{equation}
where $T$ denotes the temperature and $h_i(t)$ is the local field seen by spin $i$ at
time $t$:
\begin{equation}
h_i(t)=\sum_{j=n.n.}J_{ij}{s_j(t)}+h_0.\;\;\;\;\;\;\;\;
\label{localh1}
\end{equation}
In this expression $h_0$ represents an external magnetic field,
and the sum in the interaction term applies only to the
nearest neighbors (three for example in an hexagonal lattice).
Without loss of generality from now on we will take $h_0=0$ and $J_{ij}=1$.

\subsection{Damage spreading and self-overlap}
As stated above we will use the SO method to study the dynamics of
the system. This procedure was introduced by Luque and Ferrera
\cite{SO1} and its underlying
 philosophy is similar to that of the DS method used by Vojta to study the
thermodynamics of phase transitions in spin systems. The main
difference between the two procedures lies in that, while damage
spreading uses two copies of a system with slightly different
initial conditions (the \emph{damage}) and computes the evolution of
these differences, the self-overlap method uses the difference
between successive temporal states of a single system \emph{as the
system evolves towards equilibrium}. For instance, in DS the damage
$D(t)$ at time $t$ is defined as:

\begin{equation}
D(t)=\frac{1}{2N}\sum_{i=1}^N \bigg|\
s^{(1)}_i(t)-s^{(2)}_i(t)\bigg|
\end{equation}

and measures the (averaged) Hamming distance between the states of
the two replicas at that time (i.e., the proportion of sites for
which the spin state differs between the system $(1)$ and the
damaged replica (2)). In SO however the self-overlap $a(t)$ at time
$t$ is defined as one minus the averaged Hamming distance between
the states of a spin site at time $t-1$ and at time $t$:

\begin{equation}
a(t)= 1- \frac{1}{2N}\sum_{i=1}^N \bigg|\ s_i(t)-s_i(t-1)\bigg|.
\label{SO}
\end{equation}

%We can define too a self-distance as $d(t)=1-a(t)$.
In order to describe the time evolution of the system it is useful
to define the ``up state self-overlap'' $a_{++}(t)$ at time $t$ as
the average number of spin sites that had $s_{i}=+1$ both at time
$t-1$ and at time $t$. We also define $a_{--}(t)$, $a_{+-}(t)$ and
$a_{-+}(t)$ in a completely similar fashion. By normalization we
must then have:
\begin{equation}
a_{++}(t)+a_{--}(t)+a_{+-}(t)+a_{-+}(t)=1.  \label{anormal}
\end{equation}
Since the sites that remain in the same state at times $t-1$ and $t$
drop from the sum in the definition \eqref{SO} we also must have
\begin{equation}
a(t)=1-a_{+-}(t)-a_{-+}(t)=a_{++}(t)+a_{--}(t).  \label{a}
\end{equation}
Once the equilibrium as been reached the relation $a_{+-}=a_{-+}$
must be satisfied, where we have dropped the time dependence to
indicate equilibrium values. Then trivially
\begin{equation}
a_{+-}=a_{-+}=\frac{1-a}{2}.  \label{apm}
\end{equation}
At this point it is interesting to note that the self-overlap
functions can be understood in terms of autocorrelation functions,
more precisely, two-time autocorrelation functions. For instance, in
equation (\ref{SO}), one can rewrite $|s_i(t)-s_i(t-1)|$ as
$(1-s_i(t)s_i(t-1))/2$, which is a shifted autoresponse function
measured at subsequent time steps. In a similar way, the rest of
self-overlap functions
can be written as linear combinations of the basis of autoresponse functions.\\
Autocorrelation functions have been widely used as efficient tools
in order to measure spatial or temporal correlations in physical and
biological systems (repeated patterns, relaxation, frustration,
etc). Their applications range from investigations in transport
properties of fluids \cite{fluidos} or the analysis of
climatological models \cite{clima} to studies of decoherence in
quantum systems \cite{quantum}, to cite but a few. Autocorrelation
functions are the center of interest in theoretical studies of the
relaxation of non-equilibrium systems. In this sense, much work has
been recently done in order to characterize dynamical scaling and
other invariant behavior in the ageing regimes of Ising-like systems
\cite{AG1,AG2,AG3,AG4,AG5}. In our case it would be fair to say that
the self-overlap functions are really measurements of
autocorrelations under a different garment. To dwell on a deeper
review of the existing literature on autoresponse functions would go
beyond the scope of this paper however. We would like to emphasize
nonetheless that what is new here is: \emph{i)} the fact that this
particular combination of self-correlation functions measured at
subsequent time steps manages to capture the essence of the
(same-site) temporal correlations in systems that undergo
order/disorder phase transitions, and \emph{ii)} this is then
combined with a philosophy inspired by DS, namely: an evolution
equation towards the equilibrium state for the correlations, and a
mean field approximation directly extracted from DS in order to be
able to solve this equation. Once the evolution equation and the
mean field approximation are in place the self-overlaps will allow
us to study the stability of the different states accessible to the
system, and hence the phase transition itself.

%%%%%%%%%%%%%%%%%%%%%%%%%%%%%%%%%%%%%%%%%%%%%%%%%%%%%%%%%%%%%%%%%%%%%%%%%%%%%%%%%%%%%%%%%%
%%%%%%%%%%%%%%%%%%%%%%%%%%%%%%%%%%%%%%%%%%%%%%%%%%%%%%%%%%%%%%%%%%%%%%%%%%%%%%%%%%%%%%%%%%
\section{Master equation, transition probabilities, and mean field}
\subsection{Master equation}
 Generally speaking, the self-overlap method would proceed
by solving some master evolution equation for the $a$'s in order to
obtain their equilibrium values, much in the vein of the damage
spread method. We begin by defining the probability of finding a
spin site in the $+$ ($-$) state at time $t$, $P_{+}(t)$
($P_{-}(t)$)
\begin{equation}
P_{\pm }(t)=\frac{n_{\pm }(t)}N,  \label{P(pm)}
\end{equation}
where $n_{\pm }(t)$ is the number of sites with spin up (down) at
time $t$. Obviously $P_{+}(t)+P_{-}(t)=1$. By the definition of
$a_{++}(t),a_{-+}(t)$ it follows that
\begin{equation}
P_{+}(t)= a_{-+}(t)+a_{++}(t)={1+a_{++}(t)-a_{--}(t) \over 2}
\label{P+}
\end{equation}
and analogously for the down states
\begin{equation}
P_{-}(t)= a_{+-}(t)+a_{--}(t)={1+a_{--}(t)-a_{++}(t) \over 2}.
\label{P-}
\end{equation}
As noted above in the limit $t\rightarrow \infty $ the $a$'s ought
to reach their equilibrium values and one can drop the $t$ dependence.

Of particular interest to us will be the transition probabilities
from one state to another, i.e., the elements of the transition
matrix of our Markov process. Let $W_{++}(t)$ ($W_{--}(t)$) be the
{\it average} probability of changing from the $+$ ($-$) state at
time $t$ to the $+$ ($-$) state at time $t+1$, where the precise
meaning of this average will be made clear shortly. In an mean field
approximation we will then have
\[
a_{++}(t)=W_{++}(t-1)P_{+}(t-1),
\]
\begin{equation}
a_{--}(t)=W_{--}(t-1)P_{-}(t-1) \label{a'p's}
\end{equation}
and analogously with $W_{+-},W_{-+}$. Note that these $W$'s will
then be the elements of an average Markov matrix for the evolution
of the system. Combining (\ref{P+}),\eqref{P-} and (\ref{a'p's}) together it
is easy to arrive to a couple of mean field evolution equations for
$a_{++}(t)$ and $a_{--}(t)$, namely
\begin{eqnarray}
{\frac d{dt}}a_{++}(t) &=&-a_{++}(t)W_{+-}(t)+
a_{-+}(t)W_{++}(t){,}  \nonumber \\
{\frac d{dt}}a_{--}(t) &=&-a_{--}(t) W_{-+}(t)+a_{+-}(t)W_{--}(t).
\label{aevolve}
\end{eqnarray}
These two equations are of course nothing but the reaction-diffusion
equations for the $a$'s that common sense would have dictated us to
begin with. We now proceed to evaluate a mean field approximation
for the $W$'s so that we may solve (\ref {aevolve}).

\subsection{Mean-field approximation}

To begin with, note that in a system that follows Glauber dynamics
the transition probability at site $i$ for a given local field $h_i$
is given by (\ref{fi}) above. This means that
\begin{eqnarray}
W_{++}(h_i) &=&\phi (h_i),\qquad W_{+
-}(h_i)=1-\phi (h_i),  \nonumber \\
\;W_{-+}(h_i) &=&\phi (h_i),\qquad W_{--}(h_i)=1-\phi (h_i).
\nonumber\\ \label{p}
\end{eqnarray}
That is, as is well known for a given local field $h_i$ the
probability that the spin at site $i$ will be in the $+$ state at
time $t+1$ is always $\phi (h_i)$, whereas the probability that its
state be $-$ will be $1-\phi (h_i)$, regardless of the initial state
of the site. Thus, finding average values for the $W$'s is
equivalent to finding an average $\phi (h_i)$,
$\overline \phi $.%
%. A first, back of the envelope type, mean field approximation readily
%suggests itself: since we are interested in the behavior at equilibrium,
%simply use the value of the exact Onsager solution for the average
%magnetization $m$ as the spin per site, whence the mean field seen by all
%spins will be $h_i=4m$ and we can take $<\phi >$ to be $\phi (4m),$
%\begin{equation}
%<\phi >={\frac{e^{4m/T}}{e^{4m/T}+e^{-4m/T}}}.\;\;\;\;\;\;\;\;
%\label{averfi1}
%\end{equation}
%Taking (\ref{averfi}), (\ref{p}) and (\ref{a'p's}) into the definition of $a$
%(\ref{a}), trivially yields
%\begin{equation}
%a=\phi (4m)\frac{1+m}2+(1-\phi (4m))\frac{1-m}2,\;\;\;\;\;\;\;\;  \label{a1}
%\end{equation}
%since at equilibrium $m$ does not depend on time.%
%%%%%%%%%%%%%%%%%%%%%%%%%%%%%%%%%%%%%%%%%%%%%%%%%%%%%

The mean field approximation that we will use closely follows the
spirit of the effective-field approximation used by Vojta
\cite{Vojta1}. This consists basically in averaging over all the
possible configurations that can surround a given site, where in the average each
configuration is weighted by its probability of taking place.
Thus, with three nearest neighbors per site,
the
transition probabilities can take the values (remember that we are taking $%
J_{{ij}}=1$)
\[
\phi _0\{+++\}=\phi (3)={\frac{e^{3/T}}{2\cosh {(3/T)}}},
\]
\[
\phi _1\{++-\}=\phi (1)={\frac{e^{1/T}}{2\cosh {(1/T)}}},
\]
\[
\phi _2\{+--\}=\phi (-1)={\frac{e^{-1/T}}{2\cosh {(-1/T)}}},
\]
\begin{equation}
\phi _3\{---\}=\phi (-3)={\frac{e^{-3/T}}{2\cosh {(-3/T)}}}.
\label{phi's}
\end{equation}
Note that the calculations are much simpler than those needed in DS
\cite{Vojta1}. The probability associated to each configuration will
be
\[
P(\phi _0)=P_{+}^3,
\]
\[
P(\phi _1)=3P_{+}^2(1-P_{+}),
\]
\[
P(\phi _2)=3P_{+}(1-P_{+})^2,
\]
\begin{equation}
P(\phi _3)=(1-P_{+})^3, \label{P's}
\end{equation}
where to simplify the notation we have dropped the time dependence,
although in this case one must be aware that we are not dealing with
equilibrium values (this will be the case for the next several
equations). Using equations (\ref{P+}) and (\ref{P-}), we can now
write after some trivial manipulations
\begin{equation}
\overline{\phi}=\sum_{k=0}^3P(\phi _k)\phi _k =\nonumber
\end{equation}
\begin{eqnarray}
{\frac 12}+{\frac 38}(a_{++}-a_{--})\Biggl[\tanh {\Biggl({\frac
3T}\Biggr)}+ \tanh {\Biggl({\frac 1T}\Biggr)}\Biggl] \nonumber\\
+{1 \over 8}(a_{++}-a_{--})^3 \Biggl [ \tanh {\Biggl({\frac
3T}\Biggr)} -3\tanh {\Biggl({\frac 1T}\Biggr)}\Biggr ].
\label{fiaver2}
\end{eqnarray}

Using the relations between the $a$'s and applying the mean field to
the right hand side of the differential equations (\ref{aevolve}) we
can rewrite them as
\begin{equation}
{\frac{d }{dt}}a_{++}=-a_{++}(1-\overline{\phi}) +\left({\frac{1-
a_{++}- a_{--} }{2}}\right)\overline{\phi}\label{evolucion1}
\end{equation}
\begin{equation}
{\frac{d }{dt}}a_{--}=-a_{--}\overline{\phi} +\left({\frac{1-
a_{++}- a_{--} }{2}}\right)(1-\overline{\phi}),\label{evolucion2}
\end{equation}
which by \eqref{fiaver2} is now a system of equations depending only
on $a_{++}$ and $a_{--}$. Note that it is easy to generalize the
mean field approximation to the case of $n$ nearest neighbors (that
is, for a given topology):
\begin{equation}
\overline{\phi}=\sum_{k=0}^n {n \choose k}P_+^{n-k}P_-^k {1 \over
1+\exp \left( {2n-4k \over T} \right)}.
\end{equation}
This would be much harder to do using DS, if at all possible.

\section{Thermodynamics: magnetization}

At this point we are going to link the self-overlaps to the average
magnetization per spin, $m$. With
$P_{+}(t),P_{-}(t)$ as defined above
\begin{equation}
P_{\pm }(t)=\frac{n_{\pm }(t)}N,  \label{Ppm}
\end{equation}
we must then obviously have for the average magnetization $m$
\begin{equation}
m(t)=P_{+}(t)-P_{-}(t),  \label{mPP}
\end{equation}
or, since $P_{+}(t)+P_{-}(t)=1$,
\begin{equation}
P_{+}(t)=\frac{1+m(t)}2,\qquad \qquad P_{-}(t)=\frac{1-m(t)}2.\;
\label{Pm}
\end{equation}
By the definition of $a_{++}(t),a_{-+}(t)$ it follows then
\begin{equation}
a_{-+}(t)+a_{++}(t)=P_{+}(t)={\frac{1+m(t)}2,}  \label{a'P's}
\end{equation}
and analogously with $a_{--}(t),a_{+-}(t)$ and $P_{-}(t).$ \\
Since
\begin{equation}
m(t)=a_{++}(t)-a_{--}(t)  \label{m},
\end{equation}
the system of equations
(\ref{fiaver2},\ref{evolucion1},\ref{evolucion2}) can be rewritten
as

\[
{\frac{d }{dt}}m=\frac{m}{2}\left\{ -1+{\frac {3}{4}}
\left[\tanh\left({\frac {1}{T}}\right) +\tanh\left({\frac
{3}{T}}\right)  \right] \right\}
\]

\begin{equation}
+{m^3 \over 8} \left \{ \tanh {\left({\frac {3}{T}}\right)} -3\tanh
{\left({\frac {1}{T}}\right)} \right \}. \label{magnetizacion}
\end{equation}
Within the limits of our approximation this equation describes the
evolution towards equilibrium of the magnetization $m$ for the case
of $n=3$ nearest neighbors. Setting $dm/dt=0$ one can obtain an
expression for the temperature dependence of its equilibrium value
$m(T)$, and from it one can extract the transition temperature for
the ferro-paramagnetic transition ---this was the approach
originally followed by Vojta \cite{Vojta1}.\\
Equation (\ref{magnetizacion}) yields a critical temperature $T_c
\approx 2.104$ above which the magnetization is zero. When $T<T_c$,
we have:

\begin{equation}
m=\pm\sqrt{{  -1+{\frac {3}{4}}\left [ \tanh {\left({\frac
{3}{T}}\right)} +\tanh {\left({\frac {1}{T}}\right)}\right ]
\over{\frac {3}{4}}\tanh {\left({\frac 1T}\right)} -{\frac
{1}{4}}\tanh {\left({\frac {3}{T}}\right)}  }} \label{Tcritica}.
\end{equation}
Both results completely coincides with those in \cite{Vojta1}. Note
however that the calculations involved here have been considerably
simpler ---again basically due to the fact that in SO we only
consider one replica of the system, which results in a considerable
reduction in the number of configurations that need
to be taken into account.\\
%%%%

\begin{figure}[t]
\leavevmode \epsfxsize=7.5 cm \epsffile{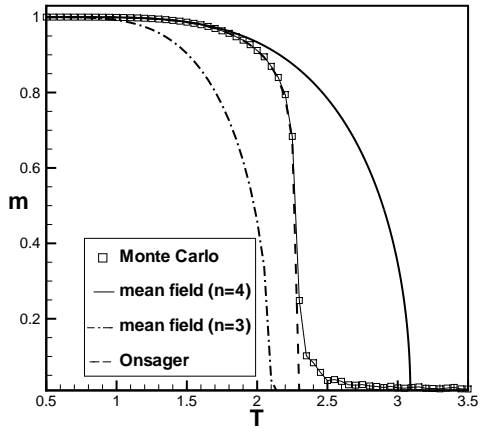}
\caption{Magnetization of the system versus temperature in the case
of: (squares) Monte Carlo simulation of a $100\times100$ spin square
lattice (the solid line here is just a guide for the eye), with
$10,000$ system steps and averaged over $100$ realizations ;
(dashed-dot line) mean field approximation for $n=3$ neighbors;
(solid line) mean field approximation for $n=4$ neighbors; (dashed
line) Onsager solution. Note that the mean field approximation
recovers the expected behavior, that is, null magnetization above
$T_c$, non null magnetization below $T_c$, which tends to a constant
maximum value at $T=0$. The difference lies on the quantitative
value of $T_c$ in each case, overestimated by the mean field in the
case of $n=4$ neighbors.} \label{mfigure}
\end{figure}

%%%%%%%
In our Monte Carlo simulations, the procedure to measure the
(equilibrium) self-overlap goes as follows: let us suppose that we
generate a random initial condition for the $N$ spin lattice. Then
we let it evolve towards equilibrium by applying the Glauber
dynamics with $4$ neighbors (square lattice). Once equilibrium has
been reached we compute the states of the system for a sufficiently
large number of time steps. We have used in all cases $10,000 \times
N$ time steps for a square lattice of $N=100\times100$ spins (that
is, defining a system time step $t$ as $N$ steps of the simulation,
we use $t=10000$ system time steps). If we then count the number of
times that a spin site is in the ``up'' state, $+$, both at time $t$
and $t-1$ and average over all sites and time steps, this will give
us the equilibrium value of the up state self-overlap $a_{++}$.
Repeating this procedure with the down state, $-$, will then
obviously give us
the ``down state self-overlap'', $a_{--}$, and so on. Each value of the simulation is averaged over $100$ realizations.\\
 In figure (\ref{mfigure}) we plot the average
equilibrium magnetization vs. temperature in order to visualize how
our mean-field approximation performs ---we note here that we are
basically interested in the thermodynamic limit of infinite lattice
size and that we are removing the inherent degeneracy of the system
by plotting only positive magnetization. First, note that our Monte
Carlo simulations in a square lattice (squares) are in fair
agreement with the Onsager (infinite size) solution --dashed line--
except in the proximity of the phase transition, where finite size
effects are relevant and difficult to suppress. Comparing then the
Monte Carlo simulations and the mean-field solution with $n=4$
neighbors we can see that qualitatively speaking they provide the
same results, with the mean field typically overestimating the
critical temperature. We stress here however that the purpose of
this paper was not so much to present a mean field technique able to
reproduce the exact results, but rather to introduce a new technique
able to exactly reproduce previously known mean field results while
at a much lower cost. For illustrative purposes and to allow
comparison with the results obtained by Vojta we also show in figure
\ref{mfigure} the mean-field result for $n=3$ neighbors (hexagonal
lattice), which underestimates the $n=4$ critical temperature $T_c$.

\begin{figure}[t]
\leavevmode \epsfxsize=7.5 cm \epsffile{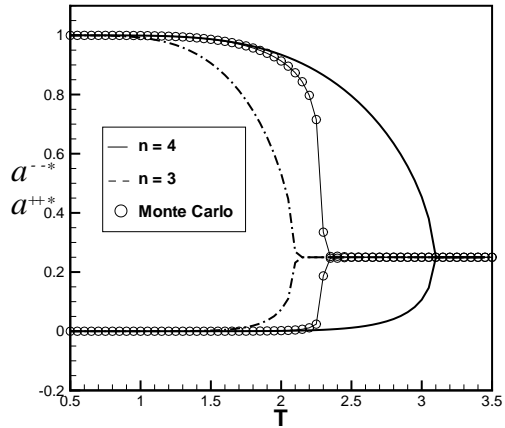} \caption{Stationary
values of $a_{++}$ and $a_{--}$ in the case of: mean-field
approximation with $n=3$ first neighbors (dashed line), mean-field
approximation with $n=4$ first neighbors (solid line) and Monte
Carlo simulation of a $100\times100$ spin square lattice (here the
solid line is just a guide for the eye), with $10,000$ system steps
and averaged over $100$ realizations (circles). Note that at $T_c$ a
pitchfork bifurcation takes place in the three cases. The
bifurcation value is underestimated by the mean field approximation
in the case of $n=3$ neighbors and overestimated in the case of
$n=4$ neighbors. Below the critical temperature $a_{++}\rightarrow
1$ while $a_{--}\rightarrow 0$ for a system that chooses the $m=+1$
vacuum, whereas the opposite is true if the system goes to $m=-1$.
Above $T_{c}$ the system tends to ($a_{++}$,$a_{--}$)=($1/4,1/4$).
Note that although the critical temperature is only predicted
qualitatively, the stationary values for ($a_{++}$,$a_{--}$) yielded
by our simple model exactly match the Onsager predictions.}
\label{fp}
\end{figure}

In figure (\ref{fp}) we plot the equilibrium values
$a_{++}^*,a_{--}^*$ vs. temperature, following the same methodology
of figure (\ref{mfigure}): we compare our Monte Carlo simulations
(circles) with the numerical resolution of the mean field equations
(note that again, the mean field with $n=3$ underestimates the
quantitative behavior and the one with $n=4$ overestimates it). As
we can see in the figure, the
self-overlap $a=a_{++}+a_{--}$ acts as an order parameter.\\%%%%%%%%%%%%%%%%%%%%%%%%%%%%%%%%%%%%%%%%%%%%%%%%

We also note that more work remains to be done in order to make an
in-depth comparison between DS and SO. For instance, one may
evaluate the critical exponents of the self-overlap order parameter
$a$ and compare the results with the DS approach \cite{DS1}, which
would be interesting. This however goes somewhat beyond the scope of
this paper.

\section{Stability}

\begin{figure}[t]
\centering \leavevmode \epsfxsize=7.5 cm \epsffile{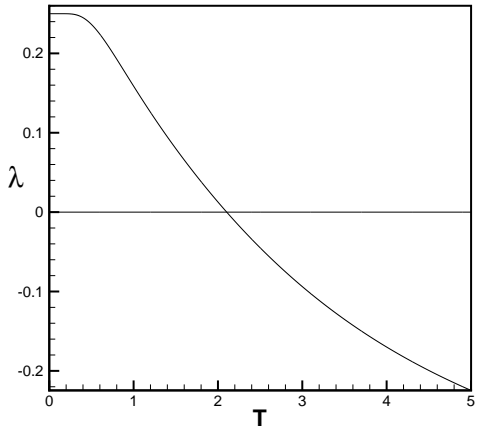}
\caption{Values of the temperature dependent eigenvalue
(\ref{lambda}) of $J$: when it is negative,
($a_{++}$,$a_{--}$)=($1/4,1/4$) is stable, thus the self-distance
$d=1/2$ is the attractor of the system (chaotic phase). When the
eigenvalue is positive, the value ($1/4,1/4$) is a saddle point and
thus an unstable fixed point. At $T_c\approx2.104$ the eigenvalue is
null, thus the fixed point is not hyperbolic -a bifurcation takes
place-.} \label{L}
\end{figure}

In equation (\ref{magnetizacion}), the stability of the fixed point
$m^*=0$ (paramagnetic phase) is related to the sign of the
eigenvalue:
\begin{equation}
\lambda(T)=1/2(-1+3/4[\tanh(1/T)+\tanh(3/T)]).\label{lambda}
\end{equation}
Note that (\ref{magnetizacion}) falls into the normal form of a
pitchfork bifurcation at $T=T_c$ where the fixed point is not
hyperbolic and the Hartman-Grobman theorem \cite{guc} does not
apply. For $T>T_c$, $m^*=0$ is stable, and below it, it becomes
unstable. The fact that we have a pitchfork bifurcation at $T_c$
implies that in the ferromagnetic phase (i.e., below $T_{c}$) two
other stable fixed points must appear. They are indeed $\pm m^*$,
where
$m^*$ is now given by (\ref{Tcritica}).\\
Taking into account the relation between $m$ and $a$, with a little
algebra we arrive at
\begin{equation}
a^*=m^*(1-\overline{\phi}^*)+\overline{\phi}^*.
\end{equation}
Hence, the fixed point $m^*=0$ leads to $\overline{\phi}^*=1/2$
(according to (\ref{fiaver2})) and $a^*=1/2$, which are thus stable
at $T>T_c$. Note that $a=1/2$ is the
minimal self-overlap that the system can show.\\
We can define at this point a Hamming-like distance between
successive temporal states (a self-distance), as:
\begin{equation}
d(t)=1-a(t).\label{selfdistance}
\end{equation}
The fixed point $a^*=1/2$ implies that we must have a fixed point
for $d$ at $d^*=1/2$ which, since it is taking place at the minimal
self-overlap, is equivalent to the maximal self-distance of the
system (total disorder).
 Following Wolf's method as in the case of random Boolean
networks \cite{SO2}, this self-distance would enable us to determine
a Lyapunov exponent of the system. However, one can simply apply the
Hartman-Grobman theorem directly \cite{guc}. Near the fixed points
the self-distance of our system can be expressed in terms of
$d(t)\sim\exp(\lambda t)$, where $\lambda$ is given by
(\ref{lambda}) . This eigenvalue can also be understood as a
Lyapunov exponent. Note that nevertheless it would not be a standard
Lyapunov exponent: when $d(t)$ tends to its fixed point,
the system is actually tending to the maximal disorder, thus $\lambda<0$ means chaos.\\
Summing up, in the paramagnetic phase, $m^*=0$ is stable, thus
$d^*=1/2$ is stable too: the system tends exponentially to the
maximal disorder and the phase is chaotic.\\
Figure (\ref{L}) is a plot of equation (\ref{lambda}).
 Note that when $T>T_c$ (paramagnetic phase) an increase
of the temperature leads to an increase of chaos, with the
self-distance of the system
tending faster to the attractor $d^*=1/2$.\\\\
In the ferromagnetic phase however the stable stationary value of
$d$ is
\begin{equation}
d^*(T)=1-m^*(T)(1-\overline{\phi}^*)-\overline{\phi}^*,
\end{equation}
with $m^*$ given by (\ref{Tcritica}) and $\overline{\phi}^*$ the
fixed point value of the mean field. The self-distance tends to zero
for low $T$, and thus the system is in a frozen state (order). When
we increase the temperature the self-distance also increases up to
the maximum value $d=1/2$, which is reached at $T_c$ (no
correlation). These results agree with those found in the
paramagnetic phase. We can conclude therefore that our approach
correctly reproduces an ordered behavior in the ferromagnetic phase
and disordered (chaotic) behavior in the paramagnetic phase. In
Appendix (A) we perform a more detailed analysis of the stability of
the system that confirms this conclusion.

\section{Conclusion}
In this paper we have introduced the self-overlap method by using it
to study both analytically and numerically the $2D$ Ising model.
Since the properties of this model are obviously well known our main
concern was to show that SO is an unambiguous method (with respect
to changes in the algorithm implementation) that correctly
reproduces the standard results while being very advantageous from
both the numerical and the analytical point of view. The SO method
could thus constitute a rather simple and efficient method of
stability analysis in this kind of multicomponent systems
(Ising-like models, spin glasses, CA, Kauffman networks, etc). Many
other physically relevant quantities in these systems (measures of
complexity, information theory measures such as the mutual
information, and so on) can be studied and measured by applying SO,
something that we think deserves further investigation. Wherever
damage spreading was supposed to have been useful and the
equilibrium state of the system is ergodic, we think that
self-overlap ought to work too and do so in a non ambiguous manner.
Moreover, it should also be more efficient numerically speaking, and
simpler from the analytical viewpoint.

\begin{acknowledgments}
We would like to thank Ignacio Parra and Jose Olarrea for their
valuable opinions and the referees for their interesting comments.
This work was funded by the Spanish Ministry of Education and
Science (Grant FIS2006-08607/ to B.L. and L.L.).
\end{acknowledgments}

\appendix

\section{Detailed analysis of the stability}
We undertake here a deeper study on the stability of the system. For
this task we go back to the evolution equations
 (\ref{evolucion1}, \ref{evolucion2}), which constitute a nonlinear
 differential system. The fixed points
of this system are obtained from equating
(\ref{evolucion1},\ref{evolucion2}) to zero (reducing the
differential system to a linear system). This yields a total of
three fixed points, namely:
\begin{eqnarray}
(a_{++}(T)^*,a_{--}(T)^*)=\bigg(\frac{\overline \phi}{2}(1+m^*),
m^*(\frac{\overline \phi}{2}-1)+\frac{\overline
\phi}{2}\bigg),\nonumber\\
(a_{++}(T)^*,a_{--}(T)^*)=\bigg(m^*(\frac{\overline
\phi}{2}-1)+\frac{\overline \phi}{2},\frac{\overline \phi}{2}(1+m^*)
\bigg),\label{puntosfijos2}
\end{eqnarray}
when $T<T_c$ (where $m^*$ is given by (\ref{Tcritica})), and
$(a_{++}(T)^*,a_{--}(T)^*)=(1/4,1/4)$  $\forall T$ (this solution is
obviously related to the fixed point $m^*=0$).\\
We can write $\overline \phi$ as
\begin{equation}
\overline{\phi}=\frac{1}{2}+
\frac{1}{2}A(T)(a_{++}-a_{--})+\frac{1}{8}B(T)(a_{++}-a_{--})^3
\label{fi2},
\end{equation}
where
\begin{equation}
A(T)=\frac{3}{4}[\tanh(3/T)+\tanh(1/T)], \label{A}
\end{equation}
and
\begin{equation}
B(T)=[\tanh(3/T)-3\tanh(1/T)]. \label{B}
\end{equation}

Let's start with the stability analysis of the fixed point
$(a_{++}^*,a_{--}^*)=(1/4,1/4)$. This solution is independent of $T$
and for $T>T_c$ is the only fixed point (note that in this case
$\overline \phi$ takes the value $1/2$ independently of the number
$n$ of neighbors as it can be proved after some trivial algebra) .
Computing the jacobian $J$ at this fixed point, we come to:\\\\
$J\mid_{(1/4,1/4)}=$$\frac{1}{4}\left(
  \begin{array}{cc}
    A(T)-3 & -A(T)-1 \\
    A(T)-1 & -A(T)-3 \\
  \end{array}
\right) $,
\\\\
with eigenvalues $\lambda_1=-1$, and $\lambda_2=1/2(A(T)-1)$. We
will distinguish then three situations: when $A(T)<1$, $(1/4,1/4)$
is an hyperbolic (indeed stable) fixed point (which is obviously
related to the fact that $m^*=0$ is stable when $T>T_c$). When
$A(T)>1$ the fixed point is again hyperbolic, but now it is unstable
(a saddle point). In these two situations we can apply the developed
formalism, due to the Hartman-Grobman theorem \cite{guc}. Hence,
$A(T)<1 \Leftrightarrow T> 2/\ln(2^{2/3}+1)\approx2.104$ (and
viceversa for $A(T)>1$).
\\\\
We thus get that when $T>T_c$ (that is, in the paramagnetic phase),
the stationary solution $(1/4,1/4)$ is stable. In the ferromagnetic
phase however ($T<T_c$) this fixed point becomes unstable.\\\\

At this point we can introduce the self-distance defined in
(\ref{selfdistance}). The stability of the $(1/4,1/4)$ solution
directly implies that $d$ will have a stable value of $1/2$ in the
paramagnetic phase, whilst this value will become unstable in the
ferromagnetic phase. Since in the paramagnetic phase $(1/4,1/4)$ is
the only fixed point the self-distance necessarily goes to the
attractor (stable fixed point) $d^*=1/2$, indeed exponentially due
to the Hartman-Grobman theorem, and the phase is thus chaotic.
However in the ferromagnetic phase $(1/4,1/4)$ is unstable: orbits
with initial conditions arbitrarily close from this fixed point will
separate from it exponentially, correlations will take place and the
phase will become ordered.\\\\
When $A(T)=1$, applying Peixoto's theorem \cite{guc}, we can
conclude that $(1/4,1/4)$ is a bifurcation point (lack of structural
stability), that is, $T_c$ constitutes a bifurcation value. What
kind of bifurcation is taking place?. It is easy to see that the
linearized system has a symmetry of the type $a_{++} - a_{--}$.
Using this symmetry, the system of equations
(\ref{evolucion1},\ref{evolucion2}) can be transformed into
(\ref{magnetizacion}). This equation falls into the normal form of a
codimension one bifurcation, a pitchfork bifurcation (indeed,
subcritical). This means that two branches of equilibria appear for
$T<T_c$ associated with values of $m\neq0$, either positive
(positive branch) or negative. Undoing the change of variables we
get that below $T_c$ we must have, for a given $T$, two extra
stationary points --other than ($1/4,1/4$)-- of the shape
[($a,b$),($b,a$)]. These fixed points correspond obviously to
(\ref{puntosfijos2}). Moreover, since as the Poincar\'e index is a
topological invariant these two new fixed points are both stable in
the ferromagnetic phase (in the paramagnetic phase the global index
is $+1$ because the fixed point $(1/4,1/4)$ is a sink, whereas in
the ferromagnetic phase $(1/4,1/4)$ is a saddle point with index
$-1$, so the other two fixed points must have index $+1$). Depending
on the initial conditions, the system will evolve to a fixed point
of the shape $(a,b)$ or to $(b,a)$. In other words, the Ising model
will give us either positive or negative magnetization in the
ferromagnetic phase, depending on the initial condition. If the
system starts at $T>T_c$, where the magnetization is zero, and we
lower its temperature below the critical one,  fluctuations will
take the system either to the upper or to the lower branch indistinctively.\\

In figure (\ref{fp}) we plot together the stationary values
$(a_{++}^*,a_{--}^*)$ of the differential system
(\ref{evolucion1},\ref{evolucion2}) for both $n=3$ and $n=4$ nearest
neighbors and the results from our Monte-Carlo simulation (again, a
square lattice of $100\times100$ spins, where we ran $10,000$ system
steps after reaching equilibrium, and averaging
 over $100$ realizations).
We can see that the results are qualitatively similar, that is, the
stationary value $(1/4,1/4)$ is stable above the Curie temperature
and unstable below it. As expected, at $T_c$ a pitchfork bifurcation
takes place and when $T<T_c$ the system has two stable fixed points,
i.e. $(a,b)$ and $(b,a)$ for each $T$.

\bibliography{apssamp}% Produces the bibliography via BibTeX.

\end{document}